\documentclass[12pt]{article}
\usepackage{epsfig}

\newcommand{\eq}{\begin{equation}}
\newcommand{\eqx}{\end{equation}}
\newcommand{\eqn}{\begin{eqnarray}}
\newcommand{\eqnx}{\end{eqnarray}}
\newcommand{\f}[2]{\frac{#1}{#2}}
\newcommand{\lm}{\lambda}

\newcommand{\tr}{\mbox{\rm tr}}
\newcommand{\al}{\alpha}

\newcommand{\dl}{\delta}
\newcommand{\Dl}{\Delta}
\renewcommand{\AA}{{\cal A}}
\newcommand{\MM}{{\cal M}}
\newcommand{\AAd}{{\cal A}^\dagger}

\newcommand{\Gam}[1]{\Gamma\left({#1}\right)}
\newcommand{\hyp}[4]{{}_2 F_1\left({#1},{#2},{#3};{#4}\right)}

\newcommand{\arr}[4]{%
\left(\begin{tabular}{cc}
$#1$ & $#2$ \\
$#3$ & $#4$
\end{tabular}\right)}

\newcommand{\rr}[4]{#1, {\it #2 \/}{\bf #3} #4}

\title{New Multicritical Random Matrix Ensembles}

\author{Romuald A. Janik\footnote{
e-mail: {\tt janik@nbi.dk}}\\ \\
The Niels Bohr Institute,\\
Blegdamsvej 17, DK-2100 Copenhagen,\\ 
Denmark\\
and\\
M. Smoluchowski Institute of Physics\\ 
Jagellonian University\\ 
Reymonta 4, 30-059 Cracow, Poland}

\begin{document}

\maketitle

\begin{abstract}
In this paper we construct a class of random matrix ensembles labelled
by a real parameter $\alpha \in (0,1)$, whose eigenvalue density near
zero behaves like $|x|^\alpha$. The eigenvalue spacing near zero scales
like $1/N^{1/(1+\alpha)}$ and thus these ensembles are representatives of
a {\em continuous} series of new universality classes.
We study these ensembles both in the bulk and on the scale of
eigenvalue spacing. In the former case we obtain formulas
for the eigenvalue density, while in the latter case
we obtain approximate expressions for the scaling functions in 
the microscopic limit using a very simple approximate method based 
on the location of zeroes of orthogonal polynomials.
\end{abstract}

\section{Introduction}

Random matrix ensembles arise in an overwhelming number of diverse
applications. 
Their utility stems from the fact that even though one 
can construct a multitude of different random matrix models, when one
studies properties on the scale of eigenvalue spacing the results
become universal and independent of the detailed structure of the
random matrix measure. Hence one can extract predictions even without
knowing the details of the appropriate microscopic model of the
phenomenon being studied. 

The type of universality class depends in general on the symmetry
properties of the random matrix model and, of course, on the
scaling properties of eigenvalue spacing. Thus one has different bulk
and edge universality regimes, as well as a {\em discrete} series of
multicritical universality regimes. Consequently only a discrete
series of (rational) scaling exponents could be obtained.

In this paper we would like to show that there exists a real {\em
continuous} range of universality regimes with generic real scaling
exponents. We will perform an explicit construction for ensembles with
eigenvalue spacing behaving like $1/N^{1/(1+\al)}$ with $\al \in (0,1)$.  

A similar {\em continuous} series of universality classes
appeared naturally in \cite{LEVY}, for random matrix
models with noncompact eigenvalue support and power-law tails in the
eigenvalue density. There the universal behaviour appeared for {\em
large} eigenvalues. In general, however, the appropriate measures
were difficult to obtain in an explicit way. This was the main
motivation for this investigation.

In this paper we would like to perform a construction of a more 
conventional class of models with compact support and where the
universal regime is located close to $\lm=0$.  

We strongly suspect that a mapping of this behaviour to infinity would
correspond to the universal scaling for the L\'evy ensembles of
\cite{LEVY}. Another possible application of similar random matrix
models (in their chiral variant) might be to model the behaviour of
low lying eigenvalues of the QCD Dirac operator {\em at} the chiral
phase transition. In this context the exponent $\alpha$ in $\rho(\lm)
\sim |\lm|^\alpha$ is the inverse of the $\delta$ critical exponent of
the chiral phase transition \cite{JV,NJL,US}. 

The plan of this paper is as follows. In section 2 we briefly recall
the main features of the old and new universality regimes in random
matrix theory. In section 3 we construct the random matrix measures of
the new multicritical ensembles and compute the eigenvalue density in
the bulk of the spectrum. In section 4 we move to a discussion of the
microscopic (presumably universal) scaling behaviour and propose a
simple approximate method based on the distribution of zeroes of
orthogonal polynomials. In section 5 we compare these approximate
formulas with exact expressions calculated numerically for
$\alpha=1/3$. We close the paper with a discussion.

\section{Universality regimes}

The universality of a correlation function in a random matrix model
means that it does not change (up to a trivial rescaling) when the
probability measure is modified. Typically the universal quantities
involve properties of the eigenvalues and their correlations
considered on the scale of eigenvalue spacing in the large $N$ limit.
For a {\em multicritical} point (see below), the modifications of the 
probability measure are restricted to those that do not destroy 
multicriticality. 
However, in general, this still leaves an infinite
dimensional space of possible deformations.       

The type of universal behaviour in random matrix models depends on the
part of the spectrum that one is studying. 
In general the scaling of eigenvalue spacing with $N$, in the vicinity
of a fixed
eigenvalue $\lm_0$, can be easily obtained by looking at the local
behaviour of the (bulk) eigenvalue density (normalized to $N$):
\eq
N \, \rho(\lm)\sim N (\lm-\lm_0)^\alpha
\eqx
Consequently the number of eigenvalues between $\lm_0$ and $\Lambda$
is approximately $n \sim N (\Lambda-\lm_0)^{\al+1}$. Reexpressing
$\Lambda$ in terms of $n$ shows that the eigenvalue spacing in the vicinity of
$\lm_0$ scales like $1/N^{1/(1+\alpha)}$.

In the bulk the eigenvalue spacing is
of the order $1/N$ --- this is the classical Wigner-Dyson regime \cite{WD}. At
the edges of the spectrum the spacing is like $1/N^{2/3}$ and one
observes universal Airy-like oscillations \cite{AIRY,BI1,DEIFT}. When
the random 
matrix posseses an additional chiral structure as in applications to QCD i.e.
\eq
\MM = \arr{0}{\AA}{\AAd}{0}
\eqx   
there appears a different universal regime close to $\lm=0$ (with eigenvalue
spacing $\sim 1/N$) \cite{CHIRAL}. 

Apart from these universal regimes, when one finetunes the potential
so that $\rho(\lm)\sim \lm^{2m}$, a discrete series of multicritical
points appear (with scaling $1/N^{1/(2m+1)}$). The behaviour on the
scale of eigenvalue spacing 
in this regime is very difficult to extract \cite{ADMN,BI2,DEIFT}. A different
(but also discrete) class of multicritical models appears when adding
an appropriate {\em fixed} matrix to the random matrix. This behaviour
and novel scaling has been analyzed both for ordinary \cite{BH} and chiral
\cite{US} random matrix models.

From the above discussion we see that in order to construct a class
of random matrix models with arbitrary real scaling exponents we have to find
a random matrix ensemble whose bulk eigenvalue spacing behaves like
\eq
\rho(\lm)\sim |\lm|^\alpha
\eqx
for real $\alpha$. We will perform the construction for $\alpha \in
(0,1)$ but an extension to other $\alpha$ should not be difficult.

\section{Eigenvalue density in the bulk}

For a hermitian random matrix model with measure
\eq
e^{-N \, \tr \, V(\MM)}
\eqx
we consider the following class of potentials
\eq
\label{e.pot}
V(\MM)=-\f{t}{p} |\MM|^p +\f{g}{2p} |\MM|^{2p}
\eqx
For hermitian matrices we may define $|\MM|^p$ in the following
way. By a unitary transformation $\MM$ can be rewritten in the form
$\MM=U \Lambda U^\dagger$ with $\Lambda=diag(\lm_1,\ldots,\lm_N)$
a diagonal matrix. Then we may define $|\MM|^p \equiv U |\Lambda|^p
U^\dagger$ where $|\Lambda|^p=diag(|\lm_1|^p,\ldots, |\lm_N|^p)$. 
In fact we only need its trace which is given by
\eq
\tr |M|^p=\sum_{i=1}^N |\lm_i|^p 
\eqx
We note that although random matrix models with potentials of the type
(\ref{e.pot}) are perfectly well defined, they seem to lack an evident
diagrammatic perturbative expansion.
Potentials with a single
power-like term were first considered in the mathematical literature
\cite{MATH} and in \cite{PASTUR}.
The parameter $p$ will be related to $\alpha$ by
\eq
p=1+\al
\eqx
In the bulk we solve for the eigenvalue density by standard saddle
point method. We are looking for a solution with a single cut
$(-a,a)$. Later we will choose $a=1$.
The expression for the Green's function is then
\eq
G(z)=\f{1}{2\pi} \sqrt{z^2-a^2} \int_{-a}^a \f{d\lm}{z-\lm}
\f{V'(\lm)}{\sqrt{a^2-\lm^2}} 
\eqx
We note that for $\alpha>0$, the derivative exists.
The cut endpoint is fixed through the constraint
\eq
\label{e.cut}
\f{1}{\pi}\int_0^a d\lm \f{\lm V'(\lm)}{\sqrt{a^2-\lm^2}}=1
\eqx
The eigenvalue density can be extracted from the imaginary part
through
\eq
\rho(z)= -\f{1}{2\pi^2}\sqrt{a^2-z^2} \cdot PV \int_{-a}^a \f{d\lm}{z-\lm}
\f{V'(\lm)}{\sqrt{a^2-\lm^2}} 
\eqx
Because of the nonanalytic power like structure of the potential
(\ref{e.pot}), it is convenient to get rid of the principal value and
rewrite the formula as an ordinary integral.
First using the fact that the potential is even we may use
\eq
\f{1}{z-\lm}-\f{1}{z+\lm}=\f{2\lm}{z^2-\lm^2}
\eqx
to obtain
\eq
\label{e.rhoin}
\rho(z)= -\f{1}{\pi^2}\sqrt{a^2-z^2} \cdot PV \int_{0}^a
\f{d\lm}{z^2-\lm^2} \f{\lm V'(\lm)}{\sqrt{a^2-\lm^2}} 
\eqx
Now we use the fact that
\eq
PV \int_{0}^a \f{d\lm}{z^2-\lm^2} \f{1}{\sqrt{a^2-\lm^2}}=0
\eqx
Therefore we may rewrite (\ref{e.rhoin}) as
\eq
\rho(z)= \f{1}{\pi^2} \sqrt{a^2-z^2} \int_{0}^a \f{z V'(z)-\lm
V'(\lm)}{z^2-\lm^2} \f{1}{\sqrt{a^2-\lm^2}}
\eqx
where we could erase the principal value as the integrand is
now nonsingular. 

We will now fix the endpoint of the cut to $a=1$. Substitution of
(\ref{e.pot}) into (\ref{e.cut}) yields
\eq
-t \f{\Gam{\f{1+p}{2}}}{\sqrt{\pi} p \Gam{\f{p}{2}}} +
g  \f{\Gam{\f{1+2p}{2}}}{\sqrt{\pi} 2p \Gam{p}} = 1
\eqx
Since we want to locate a multicritical point we have to require that
$\rho(0)=0$. This gives the second equation for $t$ and $g$:
\eq
\label{e.zerorho}
\int_0^1 \f{V'(\lm) d\lm }{\lm \sqrt{1-\lm^2}}=0
\eqx
i.e.
\eq
-t \f{\Gam{\f{p-1}{2}}}{\Gam{\f{p}{2}}} +
g  \f{\Gam{\f{2p-1}{2}}}{\Gam{p}} =0
\eqx
The solution is
\eqn
t &=& \f{2\sqrt{\pi} p \Gam{p-\f{1}{2}} \Gam{\f{p}{2}}}{%
	\Gam{p+\f{1}{2}}\Gam{\f{p-1}{2}}-
	2\Gam{p-\f{1}{2}} \Gam{\f{1+p}{2}}} \\
g &=& \f{2\sqrt{\pi} p \Gam{\f{p-1}{2}} \Gam{p}}{%
	\Gam{p+\f{1}{2}}\Gam{\f{p-1}{2}}-
	2\Gam{p-\f{1}{2}} \Gam{\f{1+p}{2}}}
\eqnx

\subsubsection*{Bulk eigenvalue density}

Once the potential is fixed let us compute explicitly the bulk
eigenvalue density
\eq
\label{e.rhobulk}
\rho(z)=\f{1}{\pi^2} \sqrt{1-z^2} \left[ -t \phi_p(z)+g
\phi_{2p}(z) \right]
\eqx
where
\eq
\phi_p(z)\equiv \int_{0}^1 \f{z^p-\lm^p}{z^2-\lm^2} \f{1}{\sqrt{1-\lm^2}}
\eqx
This integral can be explicitly expressed in terms of hypergeometric
functions
\eq
\label{e.phip}
\phi_p(z)=\f{\sqrt{\pi} \Gam{\f{p-1}{2}}}{2 \Gam{\f{p}{2}}}
\hyp{1}{1-\f{p}{2}}{\f{3-p}{2}}{z^2} +\f{\pi}{2} \tan \f{\pi p}{2}
\f{z^{p-1}}{\sqrt{1-z^2}} 
\eqx
We see that it is the last term which gives the multicritical behaviour that we
wanted to obtain $\rho(z)\sim z^{p-1}=z^\alpha$. Owing to equation
(\ref{e.zerorho}) the constant terms cancel. We could have chosen of
course a different power for the second term in (\ref{e.pot}), however with
this choice the numerical construction of the relevant orthgonal
polynomials is made easier. 

Let us note what happens when we reach the point $\alpha=1$. Then
$p=2$ and the potential obtained here is the standard multicritical
quartic one $V(x)=-4x^2+4x^4$. However due to the vanishing of the
coefficient of the last term in (\ref{e.phip}) the
eigenvalue density behaves like $\rho(z) \sim z^2$ and not $z$. When
we increase $\alpha$ the expression for the eigenvalue density ceases
to be positive and thus the construction fails. Presumably a modification
of the power in the second term of the potential might cure the
problem but we will not consider that here. To sum up, the expression
(\ref{e.rhobulk}) is nonnegative when $\al \in (0,1)$ as considered in
this paper.

\section{Behaviour near the origin --- orthogonal polynomials} 

The ultimate interest in constructing the random matrix models with the
new scaling properties is to extract universal properties which
typically occur in the microscopic regime i.e. on the scale of
eigenvalue spacing. It is well known that all the relevant properties are
encoded in the orthogonal polynomials
\eq
\int_{-\infty}^\infty dx P_n(x) P_m(x) e^{-N V(x)}=\dl_{nm} 
\eqx 
It is convenient to introduce the wavefunctions
\eq
\psi_n(x)=P_n(x) e^{-N\f{V(x)}{2}}
\eqx
Then the kernel which allows for the determination of all correlation
functions is given by the expression
\eq
\label{e.kernel}
K(z,w)=\sqrt{R_N} \f{\psi_N(z) \psi_{N-1}(w)-\psi_{N-1}(z)\psi_N(w)}{z-w}
\eqx
where $R_N$ is the recursion coefficient entering
$z\psi_n(z)=\sqrt{R_{n+1}} \psi_{n+1}(z) + \sqrt{R_{n}}
\psi_{n-1}(z)$.

The microscopic limit is defined through a rescaling
\eq
x=\f{y}{N^\f{1}{1+\al}}
\eqx
and a limit $N \to \infty$ with $y$ kept fixed. The key quantity that
determines the universal properties is the rescaled wavefunction
\eq
\lim_{N\to \infty} C_N \psi_N\left( \f{y}{N^{1/(1+\al)}} \right)=F(y)
\eqx
(as well as a similar expression with $\psi_{N-1}$), $C_N$ is a normalization
constant chosen in such a way that the limit exists. In fact the existence of
such a limit is a nontrivial and nongeneric property.

In general the determination of the scaling function $F(y)$ is a very
difficult problem c.f. \cite{ADMN,BI2,DEIFT}. One can formulate a
differential equation satisfied by the wavefunction $\psi_n(x)$, but
taking the scaling limit is extremely difficult and e.g. for the case
of quartic multicritical ensemble involves data from an auxiliary Painlev\'e
equation \cite{ADMN,BI2}. Moreover the starting point of such
considerations {\em requires} a polynomial potential, which is not the
case for our class of models. It would be interesting in the future to
explore the possibility of applying the Riemann-Hilbert methods of
\cite{DEIFT}.

Here we would like to adopt a different approach and give a very
simple but approximate method of constructing the scaling function
$F(y)$. In section 5 we will compare the approximate
solution with the numerically obtained exact result for $\alpha=1/3$,
and with the mesoscopic approximation of \cite{ADMN}.  

\subsection*{Zeroes of the orthogonal polynomials}

The starting point of our construction is the elementary fact that we may
reconstruct the orthogonal polynomial from the knowledge of its
zeroes. Since in any case we normalize the (even) wave functions through
$\psi_{2n}(0)=1$ the appropiate formula is\footnote{We always consider
even potentials.} 
\eq
\psi_{2n}(x)= e^{-N\f{V(x)}{2}} \prod_{i=1}^{n} \left(
1-\f{x^2}{\lm_i^2} \right)
\eqx

Interestingly enough there is a theorem due to Ismail \cite{MI}
which states that for an (almost) arbitrary potential $V(x)$ (see
\cite{MI}) the zeroes of the $n^{th}$ orthogonal polynomial $\lm_i$ solve
the equations
\eq
\label{e.mourad}
\sum_{1\leq k \leq n, i\neq k} \f{1}{\lm_i-\lm_k} =\f{N V'(\lm_i)}{2}
+\f{1}{2}\left(\log \f{A_n(\lm_i)}{\sqrt{R_n}}\right)' 
\eqx
i.e. these are electrostatic equilibrium positions in an external
field. The additional assumptions present in \cite{MI} serve only to
prove the uniqueness of a solution to (\ref{e.mourad}).

The correction term involves the function $A_n(x)$, given 
by the formula
\eq
\f{A_n(x)}{\sqrt{R_n}}=N\int_{-\infty}^\infty \f{V'(x)-V'(y)}{x-y}
P_n^2(y) e^{-N V(y)} dy
\eqx
For polynomial potentials of degree $m$, $A_n(x)$ is a polynomial of
at most degree $m-2$.  
Note that the above equations are exact and valid for any $N$.
We see that the last term in (\ref{e.mourad}) is suppressed w.r.t. the
ordinary potential term. 

Now in order to develop our approximation scheme let us neglect the
correction factor $(\log A_n(x))'$ (this is an approximation,
the $A_n$ term may 
indeed give some contributions in the scaling limit). 
From (\ref{e.mourad}) we see that the zeroes are distributed with a
continuum density identical to the eigenvalue density $\rho(\lm)$ of the
corresponding random matrix model. The basis of our approximation
scheme is the assumption that locally the zeroes are distributed uniformly
w.r.t. $\rho(\lm)$ (see the examples below). 

\subsubsection*{Classical Wigner-Dyson scaling functions}

Before we consider the case of immediate interest to us, in order to
illustrate the method let us 
first rederive the formulas for the standard Wigner-Dyson scaling
functions. To this end we assume that 
the eigenvalue density at $\lm=0$ is nonvanishing and $N$ is very
large. For simplicity we take the potential to be even.
It is convenient to normalize the continuum density of zeroes by
\eq
\int_{-\infty}^\infty \rho_{cont}(\lm) \, d\lm =N
\eqx
In our approximation $\rho_{cont}(\lm)$ is simply given by $N \cdot
\rho(\lm)$ where $\rho(\lm)$ is the bulk eigenvalue density of the
random matrix model (normalized to 1).
If the number of zeroes is {\em even} then they
are distributed symmetrically around 0 and there is {\em no} zero at
$\lm=0$.
The assumption of `uniform distribution' means that the $i-th$ zero is
located at $\lm_i$ where
\eq
\int_0^{\lm_i}  \rho_{cont}(\lm) d\lm =i+\f{1}{2}
\eqx
Since we are interested only in zeroes which lie close to $\lm=0$, in
the above expression we may substitute $\rho_{cont}(\lm) \to N \rho(0)$.
Consequently $\lm_i=(i+1/2)/(N\rho(0))$. Therefore the wavefunction is
\eq
\psi(x)=e^{-N\f{V(x)}{2}} \prod_{i=0}^\infty \left( 1-\left(\f{
\rho(0) xN}{i+1/2} \right)^2 \right)
\eqx
Here we extended the upper limit of the product to infinity.
After introducing the scaling variable $y=xN$ we obtain immediately
\eq
F(y)=\cos \left(\pi \rho(0) y  \right)
\eqx
the exact scaling function in the bulk. The odd case is similar but
then there is a zero at $\lm=0$ and consequently the relevant equation
is 
\eq
\int_0^{\lm_i} \rho(\lm) d\lm =i
\eqx
From the infinite product representation one obtains then the sine function.
The kernel (\ref{e.kernel}) in the scaling limit follows immediately
\eqn
K(y_1,y_2) &\sim& \f{\sin \left(\pi \rho(0) y_1 \right) \cos
\left( \pi \rho(0) y_2 \right) -
\cos \left(\pi \rho(0) y_1 \right)
\sin \left(\pi \rho(0) y_2 \right)}{y_1-y_2} \nonumber\\
&=& \f{\sin \left(\pi \rho(0) (y_1-y_2) \right)}{y_1-y_2}
\eqnx
The fact that we obtained here the {\em exact} scaling functions does
not mean, however, that we should expect to get exact results in general.

\subsubsection*{Approximate scaling function for $\al \in (0,1)$}  

The analysis of the scaling function for the new multicritical
ensembles is very similar. For 
definiteness we will just consider the even case. We start from the
behaviour of the eigenvalue density near zero:
\eq
\rho(\lm)= c_\al |\lm|^\al 
\eqx
where $c_\al$ can be easily extracted from (\ref{e.rhobulk}).
The positions of the zeroes according to the `uniform distribution'
approximation are determined by
\eq
\int_0^{\lm_i} N c_\al \lm^\al d\lm=\f{1}{\al+1}c_\al N\lm_i^{\al+1}\equiv
\f{1}{\al+1}c_\al y_i^{\al+1} =i+\f{1}{2}
\eqx
where we introduced the rescaled variable $y=\lm N^{1/(1+\al)}$.
Hence 
\eq
y_i=\left[ \f{1+\al}{c_\al}\left(i+\f{1}{2}\right) \right]^{\f{1}{1+\al}}
\eqx
Consequently the scaling function normalized by $F(0)=1$ is
\eq
\label{e.scalingf}
F(y)=\exp\left\{\f{t}{1+\al} \f{y^{1+\al}}{2}\right\} \cdot
\prod_{i=0}^\infty \left(1-\f{y^2}{y_i^2} \right) 
\eqx
In the next section we will compare this approximate result with
the exact orthogonal polynomials for $\al=1/3$.
However because the convergence properties of the infinite product are
not very good we will truncate it at some $n_{max}$ (for numerical
computation we use $n_{max}=200$) and approximate the rest of the
terms through 
\eq
\prod_{i=n_{max}+1}^N \left(1-\f{y^2}{y_i^2} \right) \sim \exp \left\{
\sum \log\left(1-\f{y^2}{y_i^2} \right) \right\} \sim
\exp \left\{ -y^2 \int_{y_{n_{max}+1}} \f{\rho(\lm)}{\lm^2} \right\}
\eqx 
The result is
\eq
\exp \left(  -\f{1}{1-\al} c_\al \f{y^2}{y_{n_{max}+1}^{1-\al}} \right)
\eqx
As we see when taking the $n_{max}\to \infty$ limit this term
becomes equal to unity (because $y_{n_{max}+1} \to \infty$). 
However we include it just because of the slow numerical
convergence of the infinite product. 

\section{An example --- $\al=\f{1}{3}$}

The orthogonal polynomials for any potential can be constructed using
the determinant formula:
\eq
P_n(x)\sim \det \left(%
\begin{tabular}{ccccc}
$\Dl_0$ & $\Dl_1$ & $\Dl_2$ & \ldots & $\Dl_n$ \\
$\Dl_1$ & $\Dl_2$ & $\Dl_3$ & \ldots & $\Dl_{n+1}$ \\ 
$\Dl_2$ & $\Dl_3$ & $\Dl_4$ & \ldots & $\Dl_{n+2}$ \\
\ldots  & \ldots  & \ldots  & \ldots & \ldots \\
$x^0$   & $x^1$   & $x^2$   & \ldots & $x^n$     
\end{tabular}
\right)
\eqx
where $\Dl_i$ are the moments
\eq
\Dl_i=\int_{-\infty}^\infty x^i e^{-NV(x)} dx
\eqx
The advantage of the specific form of (\ref{e.pot}) is that the
moments $\Dl_i$ can be expressed analytically in terms of the confluent
hypergeometric functions~${}_1F_1$.

We constructed orthogonal polynomials for $\alpha=1/3$ ($p=4/3$) for
$N=40,120$ and $N=240$. In Fig. 1 we show the wavefunctions expressed
in terms of the scaling variable $y=x N^{3/4}$
\eq
\psi_N^{sc}(y)\equiv \psi_N\left(y/N^\f{3}{4} \right)= P_N \left(y/N^\f{3}{4}
\right) \cdot e^{-\f{N}{2} V\left(y/N^\f{3}{4} \right)}
\eqx
for the above values of $N$, normalized by $\psi_N(0)=1$. We see
convergence towards a well defined scaling function.

\begin{figure}[h]
\centerline{%
\epsfysize=5cm 
\epsfbox{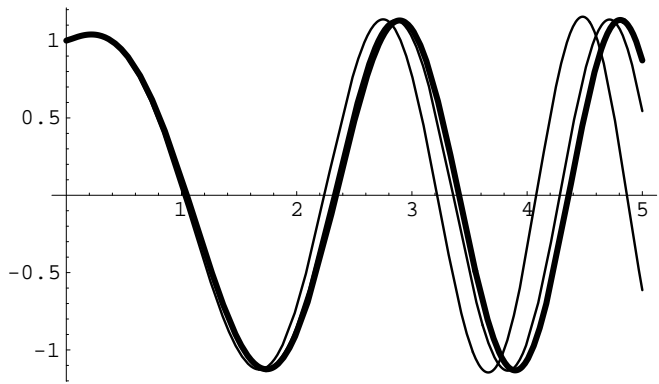}}
\caption{The wavefunctions $\psi_N^{sc}(y)$ for $\al=1/3$ expressed as a
function of the scaling variable $y$ for $N=40,120$ and $N=240$ (thick
line).}
\end{figure}

In Fig. 2 we compare the exact wavefunction for $N=240$ with the
approximate scaling function $F(y)$ obtained in the previous section. There
is no free parameter in $F(y)$. The agreement is indeed surprisingly
good. Even the small `bump' close to $y=0$ is correctly reproduced.
In fact it is difficult to judge from the numerical comparision
whether the approximation is exact or not in this case. The small
deviations might be caused either by true corrections which go beyond
our approximation or by finite size effects. The reason why such a
simple approximation scheme works so well certainly deserves further
study.

\begin{figure}[h]
\centerline{%
\epsfxsize=8cm 
\hfill \epsfbox{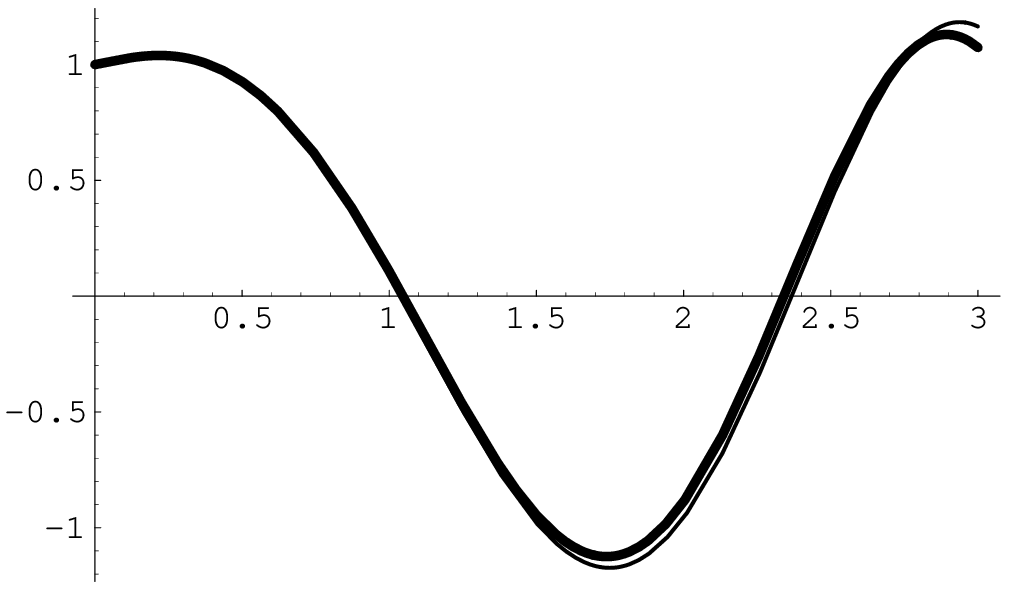} \epsfxsize=8cm 
\hfill \epsfbox{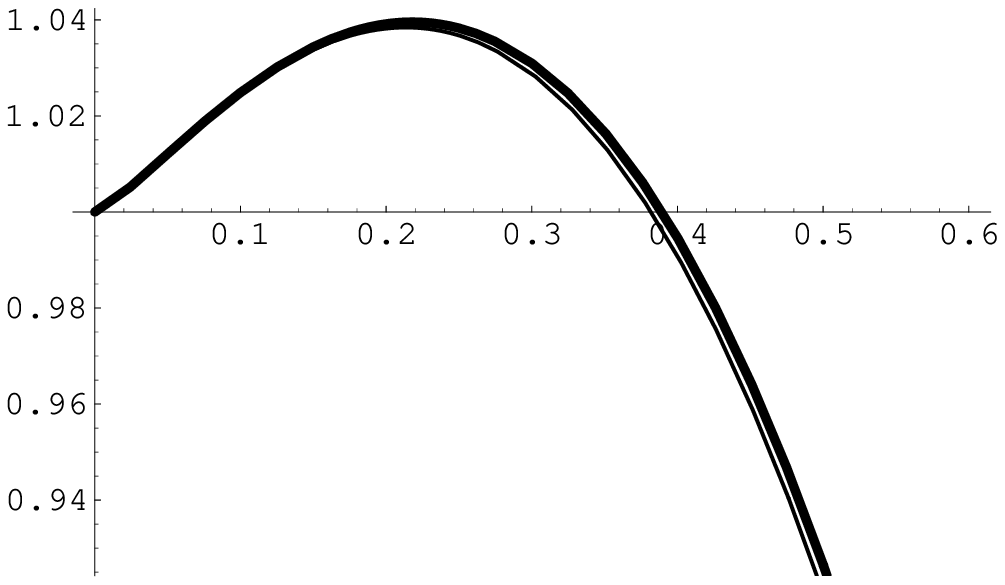} \hfill}
\caption{Comparison of the wavefunction $\psi_{240}^{sc}(y)$ (thick
line) with the approximate scaling function $F(y)$.} 
\end{figure}
  
Finally let us say a few words about the comparison with the
`mesoscopic approximation' of \cite{ADMN}. An analysis of the
differential equations for $\psi_N(x)$ for the potentials considered
here is still lacking, so we will just take a suitable analytical
continuation of the mesoscopic approximation from the discrete
multicritical points considered in \cite{ADMN} to our case. The result
is just
\eq
F_{meso}(y)=\cos \left(\f{\pi}{\al+1}c_\al y^{\al+1} \right)
\eqx
We see that the zeroes of this function {\em on the real axis}
coincide with the zeroes of our approximation, but the analytical
structure in the complex plane is certainly different. There are
spurious cuts and zeroes coming from $y^{4/3}$. Nevertheless this
approximation is also quite good but the substructure close to $y=0$ is not
captured by the mesoscopic approximation (similarly as in the quartic
multicritical case considered in \cite{ADMN}). The comparison is
presented in Fig. 3.

\begin{figure}[h]
\centerline{%
\epsfxsize=8cm 
\hfill \epsfbox{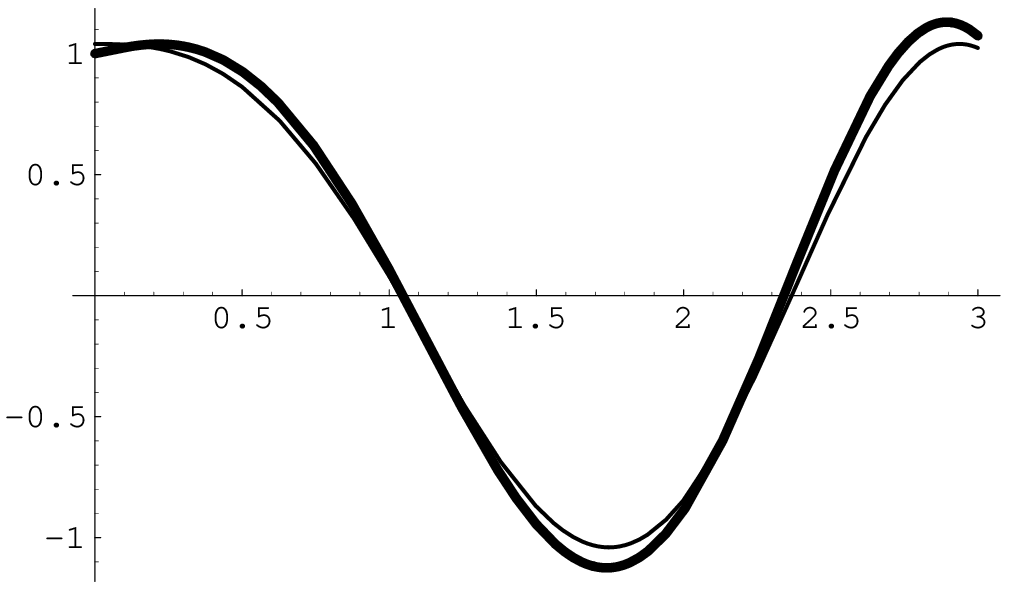} \epsfxsize=8cm
\hfill \epsfbox{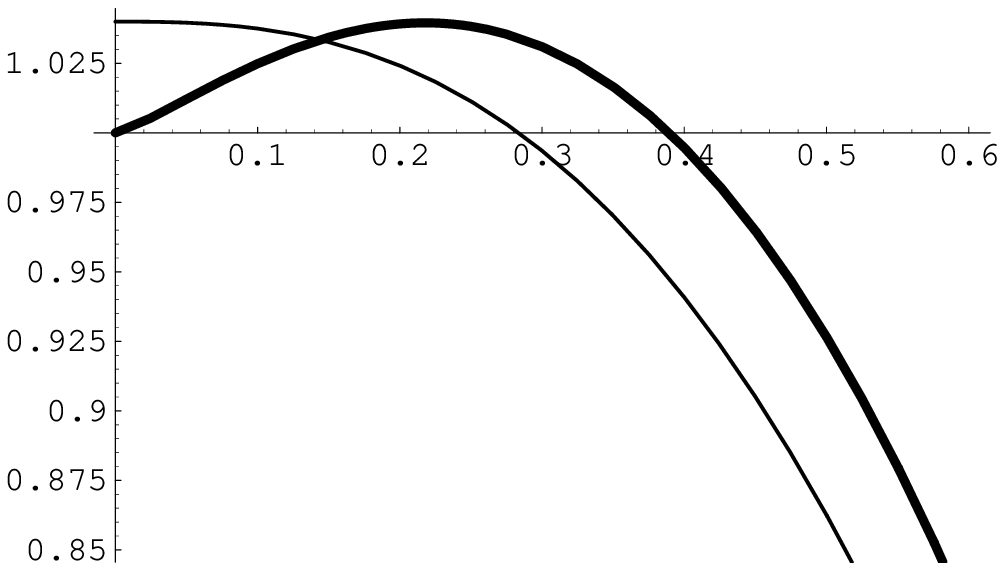} \hfill}
\caption{Comparison of the wavefunction $\psi_{240}^{sc}(y)$ (thick
line) with the mesoscopic scaling function $1.04 \cdot F_{meso}(y)$.} 
\end{figure}

\section{Discussion}

In this paper we constructed a continuous class of random matrix
ensembles which exhibit a new type of universal behaviour close to the
zero eigenvalue. The eigenvalue spacing in this region behaves like
$1/N^{1/(\al+1)}$ with a {\em real} exponent $\al \in (0,1)$.
We derived the bulk eigenvalue density and studied approximately the
behaviour of the scaling functions in the microscopic limit.
Our method for obtaining the scaling function was approximate and
based just on the local behaviour of the bulk eigenvalue
distribution. Nevertheless the approximation seems to be very good and
captures even the fine structure of the scaling functions near the origin.

It would be very interesting to see if these new universality classes
could appear in some physical systems.
We conjecture that the scaling behaviour can be directly related to
the large eigenvalue behaviour of L\'evy random matrix models
\cite{LEVY}.

There are numerous open questions and directions for further
study. Firstly it would be very interesting to try to obtain, even
implicitly through e.g. a differential equation, the exact scaling
function, especially as the vast majority of techniques dealing with
multicritical random matrix ensembles requires from the outset a 
{\em polynomial} potential. Perhaps the most promising approach would
be the Riemann-Hilbert method. On the technical side a more systematic
analysis of the properties of the orthogonal polynomial zeroes starting from
(\ref{e.mourad}), in particular the nature and scaling properties of
corrections would be very welcome. This is especially interesting in
order to better understand the effectiveness and limitiations of the
approximate method which seems to work so well here. 
In general, however, this seems to
be quite a formidable problem. 
Secondly other quantities of interest,
such as eigenvalue spacing distributions, would be important in view
of possible applications. Other possible directions are an extension
of the above considerations to chiral random matrix models (in view of
possible applications to QCD), and a
rigorous treatment of universality properties.

\bigskip
\noindent{\bf Note added:} An independent study of a very similar
class of multicritical random matrix model with real exponents 
was undertaken by Gernot Akemann and Graziano Vernizzi \cite{AV}.
\bigskip

\noindent{\bf Acknowledgments.}
I would like to thank Graziano Vernizzi for interesting discussion on
the bulk properties, Poul Damgaard for discussion on \cite{ADMN} and
Maciej A. Nowak for discussions and comments.
This work was supported in part by KBN grant 2P03B01917.

\end{document}